\documentclass[prx,aps,superscriptaddress,showpacs,floatfix,tightenlines,twocolumn]{revtex4-2}
\usepackage{graphicx}
\usepackage{dcolumn}
\usepackage{bm}
\usepackage{amssymb}
\usepackage{bbm}
\usepackage{dsfont}
\usepackage{bbold}
\usepackage{mathbbol}
\usepackage{amsmath}
\usepackage{url}
\usepackage{amsthm}
\usepackage{layout}
\usepackage{epsfig}
\usepackage{graphicx}
\usepackage{epstopdf}
\usepackage{booktabs}
\usepackage{float}
\usepackage{subfigure}
\usepackage{lipsum}
\usepackage{mathrsfs}
\usepackage{booktabs}
\usepackage{blindtext}
\usepackage{appendix}
\usepackage[hyperindex,breaklinks]{hyperref}
\usepackage{enumerate}
\usepackage{enumitem}
\usepackage{makecell}
\usepackage{booktabs}
\usepackage{multirow}
\usepackage{graphicx}
\usepackage{tabularx}
\usepackage{array}
\usepackage{subfigure}

\allowdisplaybreaks[4]
\newcommand{\ket}[1]{|#1\rangle}

\allowdisplaybreaks[3]

\usepackage{xcolor}

\newcommand{\yaqi}[1]{\textcolor{black}{#1}}

\begin{document}

\title{A scalable entanglement distribution framework for hybrid quantum networks}
\author{Yaqi Zhao}
\affiliation{{School of Mathematics, Xi’an University of Technology, Xi’an, 710054, China}}

\author{Kan He}
\email{hekan@tyut.edu.cn}
\affiliation{\yaqi{School of Mathematics and Statistics, Shaanxi Normal University, Xi’an, 710119, China}}

\author{Jinchuan Hou}
\email{jinchuanhou@aliyun.com}
\affiliation{College of Mathematics, Taiyuan University of Technology, Taiyuan, 030024, China}

\author{Yongtao~Zhang}
\affiliation{Network Science and Technology Center, Rensselaer Polytechnic Institute, Troy, New York 12180, USA}
\affiliation{Department of Physics, Applied Physics, and Astronomy, Rensselaer Polytechnic Institute, Troy, New York 12180, USA}

\author{Nicolò~Lo~Piparo}
\affiliation{Okinawa Institute of Science and Technology Graduate University,
1919-1 Tancha, Onna-son, Okinawa 904-0495, Japan}

\author{Xiangyi Meng}
\email{xmenggroup@gmail.com}
\affiliation{Network Science and Technology Center, Rensselaer Polytechnic Institute, Troy, New York 12180, USA}
\affiliation{Department of Physics, Applied Physics, and Astronomy, Rensselaer Polytechnic Institute, Troy, New York 12180, USA}
\affiliation{National Institute for Theory and Mathematics in Biology, Northwestern University and the University of Chicago, Chicago, Illinois 60611, USA}

\date{\today }
\begin{abstract} 
The realization of a scalable quantum network (QN) hinges on the efficient distribution of entanglement across heterogeneous hardware platforms. While traditional QN frameworks are often restricted to single-encoding schemes, relying exclusively on either discrete (DV) or continuous (CV) variables, such segregated architectures limit the capacity and interoperability of large-scale systems.
Here, we propose a scalable hybrid entanglement distribution scheme that bridges these domains by mapping hybrid entanglement swapping and concentration at DV--CV interfaces onto a unified set of series and parallel graph rules. We demonstrate that our hybrid approach exhibits a {potential} performance advantage over traditional point-to-point generation of maximally entangled states, achieving  enhanced end-to-end entanglement within series-parallel topologies. 
These results provide a theoretical foundation for designing resource-efficient, interpretable architectures for the future quantum internet.

\end{abstract}

\maketitle

\section{Introduction}
{Quantum entanglement is a central resource for quantum information processing~\cite{Michail2002,Michail2007,MI2000}, and its reliable distribution is a fundamental prerequisite for practical long-distance quantum communication~\cite{Gisin2007}. In a quantum network (QN)~\cite{q-internet_k08}, spatially separated nodes are limited to local operations combined with classical communication (LOCC). To overcome this locality constraint and distribute entanglement across large distances, intermediate nodes must operate as quantum repeaters. These repeaters perform key LOCC-based protocols---including entanglement swapping~\cite{MAMA1993,SVP1999}, entanglement concentration~\cite{CHSB1996,FM2001}, and more advanced distillation techniques~\cite{HWJP1998,NCHN2011,DES2017}---to create a globally coordinated \emph{scheme} for entanglement distribution.}

{Entanglement distribution has historically been developed along two parallel directions: discrete-variable (DV) and continuous-variable (CV) platforms, each offering complementary strengths. DV systems---such as qubits, which access only a finite portion of a state’s spectrum---provide high fidelity and robustness to noise but suffer from low transmission efficiency over long distances~\cite{DV_swapping_experiment1998}. CV systems---such as optical modes, which exploit the full quadrature spectrum---enable more efficient transmission but are considerably more vulnerable to noise and loss~\cite{CV_Gaussian_concentration2002,CV_concentration2006,CVrepeater2011,CV_concentration2012,CV_swapping2013}. 
These contrasting features motivate the development of \emph{hybrid} architectures~\cite{hybrid_2008,H_repeater2010,hybrid_Loock2011,hybrid_QN2013,hybrid_QN2017,hybrid_2019,hybrid_QN2019,hybrid_QN2022,CV_DV_QKD2022,hybrid_2024,hybrid2025}, which intentionally combine both encodings.
Such hybrid approaches are increasingly viewed as essential for a heterogeneous quantum internet, enabling interoperability across diverse quantum hardware platforms~\cite{hybrid_QIP2011,hybrid_2015,SMPA2015,hybrid_QN2020}.}

While recent proof-of-concept experiments have already demonstrated that robust entanglement distribution is achievable in small-scale hybrid QN settings~\cite{SMPA2015, hybrid_QN2020}, 
building a quantum internet ultimately requires entanglement distribution schemes that function across \emph{larger} scales. Significant theoretical progress has been made in DV-based QNs~\cite{Percolation_CEP2007, percolation2008,percolation2009,MJ2009,Percolation_mixed2010,SD2010,LS2011,XJS2021,OXSGBJ2022,Percolation_review2023,XYJSA2023,path-percolation_mhrk24,path-percolation_UVflower2024,Percolation_QMemo2024} and in CV-based architectures~\cite{NegPT}. However, extending the advances to a unified framework for a large hybrid QN remains an open challenge.

In this work, we establish a theoretical framework for scalable entanglement distribution in hybrid QNs.
Specifically, the QN consists of two classes of pure-state entanglement:
(i) DV two-qubit states in the form of $|\psi_{\lambda}\rangle=\sqrt{\lambda}|01\rangle+\sqrt{1-\lambda}|10\rangle$; (ii) CV two-mode squeezed vacuum states (TMSVSs) in the form of $|\psi^{{\chi}}\rangle=\sqrt{1-\chi^2}\sum_{n}\chi^n|nn\rangle$ with {the squeezing parameter $r=\rm{arctanh}\ \chi$}.
To enable entanglement distribution across such hybrid QNs, we first introduce a central organizing principle, which we term the \emph{\yaqi{homogeneous-first reduction}} strategy.
\yaqi{The key idea is to recursively extract from a hybrid QN a subgraph that is homogeneous in its resource type---either purely CV or purely DV---and reduce it using only the corresponding CV or DV protocols.
This strategy removes all applicable homogeneous series and parallel structures, leaving a reduced effective network whose remaining DV--CV interfaces are to be treated using hybrid protocols.}

\yaqi{Building on this effective network, we further introduce a scalable entanglement distribution scheme that applies hybrid entanglement swapping and concentration protocols at the remaining DV--CV interfaces.}
We adopt the total singlet conversion probability (SCP), defined as the probability of end-to-end distribution of singlet (maximally entangled state), as the performance metric for quantifying entanglement distribution efficiency. 
For general series-parallel topologies composed solely of DV--CV interfaces, we prove that our hybrid scheme {shows advantages over} the standard benchmark of pairwise (point-to-point) singlet conversion (PSC), which requires every entangled pair to be locally converted to a singlet before distribution.
This result establishes the fundamental efficiency of our framework for a broad class of network topologies.
However, for non-series-parallel architectures, such as the honeycomb lattice, the scalable hybrid protocol provides no enhancement over PSC. 
{Finally, although our analysis assumes ideal conditions, it provides a baseline for estimating experimental resources. We discuss how realistic hardware imperfections can be approximately absorbed into effective resource parameters without altering the underlying network-level transmission scheme, ultimately enabling efficient resource estimation at larger scales.}

The rest of the paper is structured as follows. In Sec.~\ref{sec-pre}, we introduce the fundamental concepts and notations. In Sec.~\ref{sec-hybridQN}, we present the general reduction strategy of hybrid QNs. In Sec.~\ref{sec-HybridSwapping}, we introduce a hybrid swapping protocol for series configurations, and in Sec.~\ref{sec-HybridConcentration} a hybrid concentration protocol for parallel settings. In Sec.~\ref{sec-DidstributionScheme}, we develop the scalable hybrid entanglement distribution scheme,  demonstrating that it enhances distribution efficiency in specific network topologies. {In Sec.~\ref{sec-implementation}, we discuss the interpretation of our findings in terms of resource estimation.} In Sec.~\ref{sec-conclusion}, a summary of our work and an outlook on several open questions are presented.

\section{Preliminaries}\label{sec-pre}
\emph{Continuous-variable (CV) and discrete-variable (DV) systems.---}A continuous-variable (CV) system is characterized by an infinite-dimensional Hilbert space and observables with continuous spectra, such as the quadrature operators of an optical mode~\cite{SP2005,Andersen2010,Holevo2012}. In contrast, discrete-variable (DV) systems operate within a finite-dimensional subspace and are described by observables with discrete eigenvalues, as in qubit or photon-number encodings~\cite{MI2000}. 
\yaqi{Homodyne detection provides a standard approach to quadrature measurements on CV modes~\cite{SP2005}, while photon counting is commonly used for DV measurements~\cite{Flamini2018}.}

\emph{Bipartite states.---}Any pure bipartite state admits a Schmidt decomposition $|\psi\rangle=\sum_n \sqrt{\lambda_n}\,|i_n j_n\rangle$, $\sum_n \lambda_n =1$,
where $\{\lambda_n\}$ are the Schmidt coefficients in the bases $\{|i_n\rangle\}_n$ and $\{|j_n\rangle\}_n$.

\emph{DV bipartite states.---}In DV systems, we focus on two-qubit pure states, which can always be written up to local unitaries as
\begin{eqnarray}\label{eq-two_qubit}
    |\psi_{\lambda}\rangle:=\sqrt{\lambda}|01\rangle+\sqrt{1-\lambda}|10\rangle,\quad \yaqi{\frac{1}{2}}\leq\lambda\leq1.
\end{eqnarray}
Their entanglement is quantified by the concurrence ~\cite{Hill1997}, $c(|\psi_\lambda\rangle)=2\sqrt{\lambda(1-\lambda)}$ given the single Schmidt coefficient $\lambda$ (always indicated in the subscript in the following). 
{Note that the state is equivalent to 
$\sqrt{\lambda}\ket{00}+\sqrt{1-\lambda}\ket{11}$ by applying a local unitary transform $\sigma_x=|0\rangle\langle1|+|1\rangle\langle0|$ at the second qubit.}

\emph{CV bipartite states.---}In CV systems, Gaussian states provide a natural description of optical modes~\cite{SP2005,CSRNTJS2012}. A typical CV state is the two-mode squeezed vacuum state (TMSVS),
  \begin{eqnarray}
    \label{eq-TMSVS}
    |\psi^{{\chi}}\rangle=\sqrt{1-\chi^{2}}\sum_{n=0}^{+\infty}\chi^{n}|nn\rangle,\quad\chi=\tanh{r},
  \end{eqnarray}
with the squeezing parameter $r>0$. Any pure bipartite Gaussian state is locally equivalent to a tensor product of TMSVSs~\cite{AR}.
The entanglement of $|\psi^{{\chi}}\rangle$ can be characterized by the ratio negativity~\cite{Ratio_negativity2024}, $\chi_{\mathcal{N}}(|\psi^{{\chi}}\rangle)=\chi$. 

\begin{figure}[]
	\centering
	\subfigure[\ Simple series QN.]{
       \includegraphics[width=1.59in]{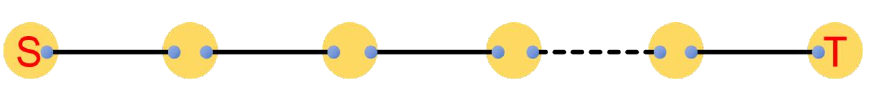} \label{fig-series}
    }
    \subfigure[\ Simple parallel QN.]{
       \includegraphics[width=1.59in]{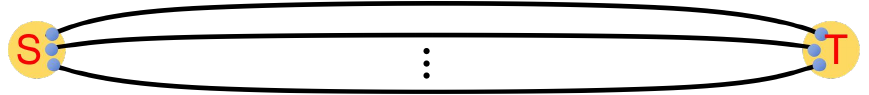}\label{fig-parallel}
    }\vspace{-2mm}
    \quad
    \subfigure[\ Parallel-then-series QN.]{
       \includegraphics[width=1.59in]{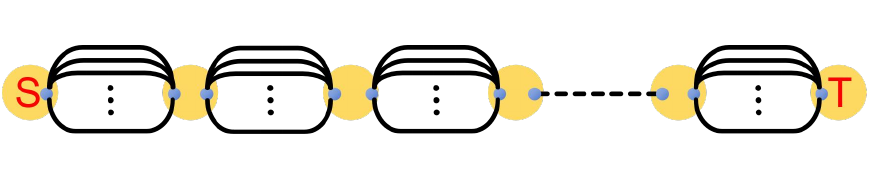}\label{fig-p-t-s}
    }
    \subfigure[\ Series-then-parallel QN.]{
       \includegraphics[width=1.59in]{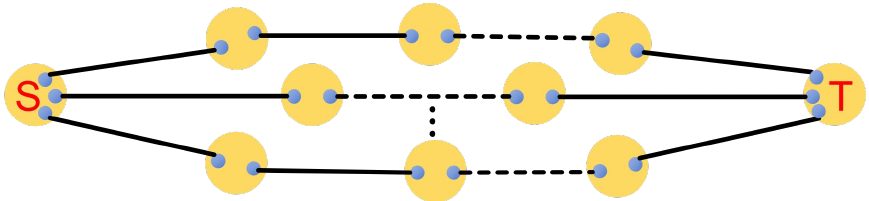} \label{fig-s-t-p}
    }\vspace{-2mm}
    \quad
    \subfigure[\ Series-parallel QN.]{
       \includegraphics[width=1.59in]{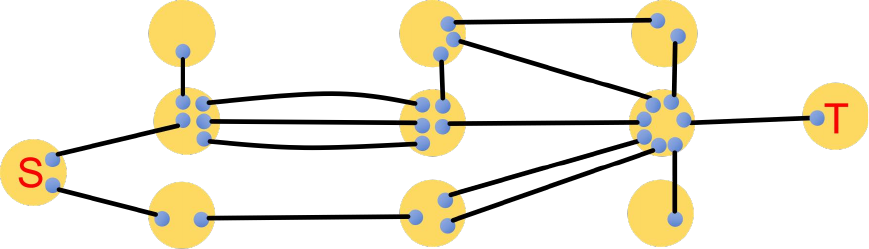}\label{fig-s-p}
    }
    \subfigure[\ Non-series-parallel QN.]{
       \includegraphics[width=1.59in]{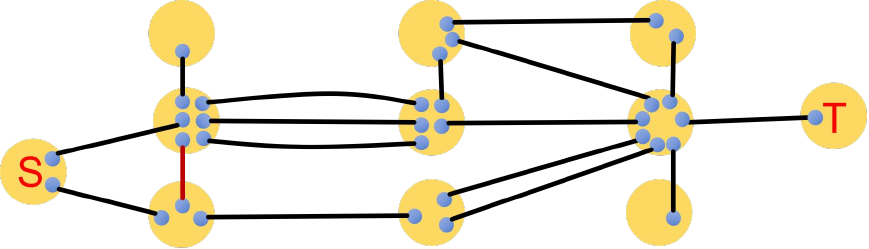}\label{fig-general}
    }\vspace{-2mm}
	\caption{End-to-end  quantum network (QN) topology. The topology between a source node $S$ and a target node $T$ can either be \subref{fig-series}--\subref{fig-s-p}~series-parallel or \subref{fig-general}~non-series-parallel. 
    }
	\label{fig-series_parallel}
\end{figure}

\begin{center}
    \begin{table*}[]
    \caption{Deterministic entanglement transmission schemes.}
    \label{table_det}
    \begin{tabular}{c|c|c|c|c}
		\hline\hline
		 Source & Scheme (Measure)  & {Initial entanglement}  & {Series rule} & {Parallel rule}\\
		\hline
       DV (two-qubit) & DET (Concurrence $c$) & $c_1,c_2,...$ & $c=c_{1}c_{2}\cdots$ & $\frac{1+\sqrt{1-c^2}}{2}=\max\left\{\frac{1}{2},\frac{1+\sqrt{1-c_1^2}}{2}\frac{1+\sqrt{1-c_2^2}}{2}\cdots\right\}$ \\
		CV (TMSVS) & G-G DET (Ratio negativity $\chi$) & $\chi_1\geq\chi_2\geq\cdots$ &  $\chi=\chi_1\chi_2\cdots$  & $\frac{\chi^2}{1-\chi^2}=\frac{\chi_1^2}{(1-\chi_1^2)(1-\chi_2^2)\cdots}$\\
		\hline\hline
	\end{tabular}
\end{table*}
\end{center}

\emph{Majorization theory~\cite{Matrix_analysis_Bhatia,VG2010} and its application for state conversions~\cite{Nielsen1999,MRN}.---}
Let $\bm x$ and $\bm y$ be two vectors of non-negative real numbers, and let ${\bm x}^\downarrow=(x_1^\downarrow,x_2^\downarrow,...)$ and ${\bm y}^\downarrow=(y_1^\downarrow,y_2^\downarrow,...)$ {denote the vectors obtained by sorting the entries of $\bm x$ and $\bm y$ in non-increasing order.}
The vector $\bm x$ is said to be majorized by $\bm y$, written ${\bm x}\prec {\bm y}$, if 
$\sum_{j}x_{j}^{\downarrow}=\sum_{j}y_{j}^{\downarrow}$ and the inequality 
$\sum_{j=1}^{k}x_{j}^{\downarrow}\leq\sum_{j=1}^{k}y_{j}^{\downarrow}$ holds for all $k$.

A bipartite state $|\phi\rangle$ can be deterministically transformed into another one $|\psi\rangle$ by LOCC if and only if the two vectors ${\bm{\lambda}}_{\phi}$ and ${\bm{\lambda}}_{\psi}$ consisting of Schmidt coefficients of $|\phi\rangle$ and $|\psi\rangle$, respectively, satisfy the majorization relation ${\bm{\lambda}}_{\phi}\prec{\bm{\lambda}}_{\psi}$~\cite{Nielsen1999,MRN}.

\emph{Singlet conversion probability (SCP)~\cite{Percolation_CEP2007}.---}Given a pure bipartite state $|\phi\rangle$ with Schmidt coefficients $\{\lambda_n\}_n$ satisfying $\lambda_1\geq\lambda_2\geq\dots\geq\lambda_n>0$.
The SCP is defined as the maximal probability to convert $|\phi\rangle$ into a singlet $\yaqi{|\Psi^-\rangle}=\sqrt{1/2}|01\rangle-\sqrt{1/2}|10\rangle$ under local operations and classical communication (LOCC), 
\begin{eqnarray}\label{eq-SCP}
    P_\text{SCP}(|\phi\rangle)=\min\{1,2(1-\lambda_1)\}.
\end{eqnarray}

\emph{Entanglement distribution schemes for QNs.---}We consider a QN in which nodes represent parties and links correspond to shared bipartite entangled states. In series-parallel networks [Figs.~\ref{fig-series}--\ref{fig-s-p}]~\cite{Duffin1965}, entanglement distribution between two boundaries (or end nodes) is carried out through two LOCC-based {protocols}:
\begin{enumerate}
[label=(\arabic*),itemsep=0pt,topsep=0pt,parsep=0pt]
    \item \emph{Entanglement swapping}---applied to links arranged in series [Fig.~\ref{fig-series}], converting multiple intermediate links into a single end-to-end entangled state.
    \item \emph{Entanglement concentration}---applied to links arranged in parallel [Fig.~\ref{fig-parallel}], combining several shared entangled states into a single state with higher entanglement.
\end{enumerate}
The entanglement swapping and concentration protocols serve as the quantum analogues to the series and parallel rules, respectively~\cite{XJS2021,NegPT}. By applying these rules, a complex series-parallel network can be simplified through a reduction process similar to the classic techniques used in circuit theory. This allows for the systematic transformation of intricate topologies into manageable, effective links.

Two primary entanglement distribution schemes based on the series and parallel rules are well established: the deterministic entanglement transmission (DET) scheme for DV-based QNs with pure two-qubit states~\cite{XJS2021}, and the Gaussian-to-Gaussian (G-G) DET scheme for CV-based QNs with TMSVS resources~\cite{NegPT} (see Table~\ref{table_det}).
In both cases, the resulting end-to-end \yaqi{states retain} the same functional form as the input resources: a pure two-qubit state for DV systems and a TMSVS for CV systems.

For networks that are not series-parallel [Fig.~\ref{fig-general}], practical entanglement distribution may be approximated using the star-mesh transform, a standard tool from graph theory that reduces general network structures into effective series-parallel equivalents.

\yaqi{
\section{Homogeneous-first reduction of hybrid QN}}
\label{sec-hybridQN}

{In practical QN settings, CV resources are well suited for short-distance metropolitan links, as they can be generated at high rates and distributed efficiently through optical channels. However, their strong sensitivity to loss and noise limits their use over long distances. DV entangled states, although typically more demanding to produce and multiplex, provide significantly higher robustness and are therefore preferred for long-haul connections~\cite{CV_DV_QKD2022}. A hybrid QN integrates these complementary capabilities, combining the high-rate distribution achievable with CV links and the long-distance resilience of DV resources, thereby overcoming the distance-efficiency trade-offs inherent to homogeneous architectures.}

\begin{figure}[]
    \centering
    \includegraphics[height=80pt]{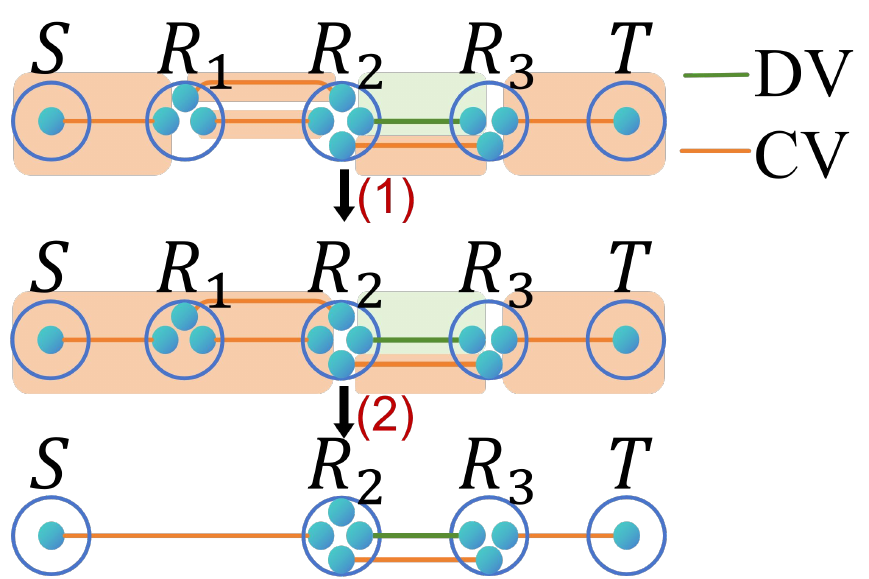}
    \vspace{-2mm}
    \caption{{\yaqi{Homogeneous-first reduction} of a hybrid QN. The network architecture comprises a mixture of {CV links} (TMSVSs, orange) and {DV links} (two-qubit states, green). {(Step 1)} Locally homogeneous subgraphs consisting of only a single resource type are repeatedly identified within the intertwined network, such as the $\{S,R_1,R_2\}$ and $\{R_3,T\}$ clusters. {(Step 2)} Each identified subgraph is independently reduced to a single effective link following the resource-specific series and parallel rules specified in Table~\ref{table_det}. This reduction transforms the complex initial topology into an {irreducible hybrid QN}, where no further homogeneous reduction rules remain applicable.}
    } \label{fig-hybridQN}
\end{figure}

We consider a hybrid QN comprising a mixture of CV  and DV links. In general architectures, these links may be interconnected in complex arrangements. To manage this complexity, we employ a \yaqi{homogeneous-first reduction} strategy to preprocess the QN (Fig.~\ref{fig-hybridQN}). Specifically, the strategy is defined by the exhaustive application of CV- and DV-specific series and parallel rules (Table~\ref{table_det}) to any applicable sub-graphs of the network. Rather than treating the network as a monolithic hybrid structure, we first identify and reduce local homogeneity: any sequence of links of the same type connected in series, or any set of links of the same type connected in parallel, is replaced by a single effective link of that type using the G-G DET or standard DET schemes, respectively~\cite{NegPT, XJS2021}.
The immediate advantage of this strategy is that both DET and G-G DET are deterministic schemes. Therefore, the network topology is deterministically reduced without the need to consider probabilistic branchings.

This reduction process continues until the network reaches an \emph{irreducible state}. In this state, no two adjacent links (in series) and no two concurrent links (in parallel) share the same resource type. The resulting network is thus ``genuinely hybrid,'' where every remaining connection represents a fundamental DV--CV interface. This strategy effectively partitions the entanglement distribution problem: the well-understood task of reducing homogeneous segments~\cite{NegPT, XJS2021} is completed first, leaving only the irreducible hybrid junctions that require new theoretical treatment.

The fundamental challenge then becomes: once this irreducible hybrid limit is reached, how should the remaining DV--CV connections be processed? Without a systematic framework for these interfaces, the end-to-end entanglement remains indeterminate. The following sections are therefore devoted to developing effective schemes for entanglement transmission within these {genuinely} hybrid topologies.

\section{Genuine Hybrid Swapping}\label{sec-HybridSwapping}



\yaqi{To establish a scalable entanglement distribution scheme based on pure-state transformations, we develop a majorization-based hybrid swapping protocol that can deterministically produce a pure two-qubit state.
Note that although there are other existing hybrid swapping protocols that also enable entanglement transfer across DV--CV interfaces, they generally produce mixed two-qubit states~\cite{RT1999,Takeda2013_1,Takeda2013_2,SMPA2015} (see Appendix~\ref{app-homodyne}). Using these outputs within a pure-state framework would therefore require additional purification. Hence, here we focus on using the majorization theory, which provides a theoretical framework for our swapping protocol.}

We begin by analyzing the elementary swapping operation for the simplest hybrid configuration of a DV and a CV link in series:
$|\psi_{\lambda}\rangle_{S R_1}\otimes|\psi^{{\chi}}\rangle_{R_1 T}$.
Here, nodes $S$ and $R_1$ share the DV state $|\psi_{\lambda}\rangle$ {[Eq.~\eqref{eq-two_qubit}]}, while nodes $R_1$ and $T$ share the CV state $|\psi^{{\chi}}\rangle$ {[Eq.~\eqref{eq-TMSVS}]}.
The hybrid swapping operation involves first converting the CV state into a DV state via LOCC,
followed by standard DV-based entanglement swapping. This method relies on an LOCC protocol~\cite{G2_conversion2025} that deterministically converts the TMSVS $|\psi^{{\chi}}\rangle$ to a two-qubit state $|\psi_{\lambda}\rangle$ for some $\lambda\geq1/2$.
Specifically, by applying the majorization criteria for state conversion detailed in Sec.~\ref{sec-pre}, one can show that the TMSVS $|\psi^{{\chi}}\rangle$ can be deterministically transformed into the following two-qubit state (up to local unitary equivalence):
\begin{eqnarray*}
      |\psi_{\eta}\rangle=\sqrt{\eta}|01\rangle+\sqrt{1-\eta}|10\rangle,
\end{eqnarray*}
where $\eta=\max\left\{{1}/{2}, 1-\chi^2 \right\}$~\cite{G2_conversion2025}.
The hybrid state $|\psi_{\lambda}\rangle_{SR}\otimes|\psi^{{\chi}}\rangle_{RT}$ thus becomes a DV state $|\psi_{\lambda}\rangle_{SR}\otimes|\psi_{\eta}\rangle_{RT}$. 

{We then implement the XZ swapping~\cite{percolation2008}. This operation achieves the optimal average concurrence through a {Bell measurement} in the $XZ$ basis.
The measurement outcomes are four distinct two-qubit states that nevertheless share identical Schmidt coefficients corresponding to $|\psi_{\xi}\rangle$, with
\begin{eqnarray}\label{eq-maj_xi}
    \xi=\frac{1}{2}\left[{1+\sqrt{1-16\eta(1-\eta)\lambda(1-\lambda)}}\right]
\end{eqnarray}
Hence, this \yaqi{measurement} enables a deterministic entanglement swapping which converts $|\psi_{\lambda}\rangle_{SR}\otimes|\psi_{\eta}\rangle_{RT}$ to a new two-qubit state $|\psi_{\xi}\rangle_{ST}$.}

{Furthermore, the resulting state {$|\psi_{\xi}\rangle_{ST}$}, which is a deterministically produced pure state but is only partially entangled, can be converted to a singlet by {a probabilistic LOCC~\cite{SCP1999}: first, perform the local measurement $\mathcal{M}=\{M_1,M_2\}$ at node $S$, where $M_1=\sqrt{\xi^{-1}(1-\xi)}|0\rangle\langle0|+|1\rangle\langle1|$, $M_2=\sqrt{\xi^{-1}(2\xi-1)}|0\rangle\langle0|$ with the outcomes labeled $m=1$ and $m=2$, respectively. When $m=1$, the LOCC is considered successful: the state $|\psi_{\xi}\rangle_{ST}$ is transformed into the maximally entangled state $|\psi_{1/2}\rangle_{ST}$---which can then be converted to a singlet by local unitaries---with the probability $P_{\text{SCP}}(|\psi_{\xi}\rangle_{ST})$ [Eq.~\eqref{eq-SCP}].}} 
These steps constitute the \emph{{majorization-based hybrid operation}}, yielding the total SCP
\begin{eqnarray*}
    p^{\text{major}}=P_{\text{SCP}}(|\psi_{\xi}\rangle_{ST})=2\left(1-\xi\right).
\end{eqnarray*}



\begin{figure}[t]
    \centering
    \includegraphics[width=220pt]{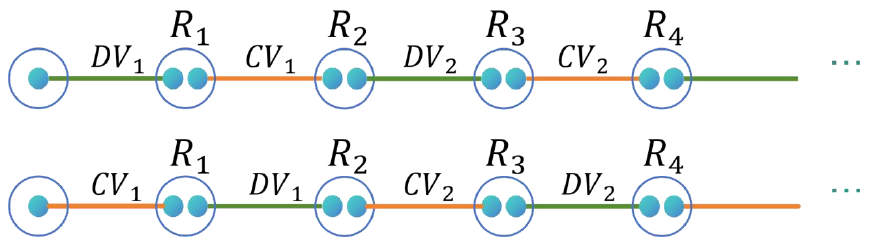}\vspace{-2mm}
    \caption{Two types of series QNs that are genuinely hybrid. The first one is DV-\yaqi{first} type with states ordered as DV, CV, DV, CV,..., and another one is CV-\yaqi{first} type, following the sequence CV, DV, CV, DV, ....
    }
    \label{fig-hybridQN4}
\end{figure}

\emph{Scalability.---}We now consider a larger, genuinely hybrid series QN obtained via the \yaqi{homogeneous-first reduction} strategy described in Sec.~\ref{sec-hybridQN}. Because the resulting network is irreducible, no two adjacent links can share the same resource type. Consequently, the network must necessarily alternate between DV and CV links.
We denote the sequence of nodes as $S, R_1, R_2, \dots,R_N, T$.
For an $N=1$ three-node chain (or any network with an odd number of nodes), the \yaqi{CV--DV} configuration $|\psi^{{\chi}}\rangle_{S R_1}\otimes|\psi_{\lambda}\rangle_{R_1 T}$ is topologically symmetric to \yaqi{the DV--CV configuration} $|\psi_{\lambda}\rangle_{S R_1}\otimes|\psi^{{\chi}}\rangle_{R_1 T}$ due to the reflection symmetry of the path. However, for a network with an even number of nodes, \yaqi{the} symmetry \yaqi{between the configurations CV--DV--CV--DV--$\dots$--CV and DV--CV--DV--CV--$\dots$--DV} is broken. \yaqi{Therefore, we} distinguish between two nonequivalent configurations (see Fig.~\ref{fig-hybridQN4}):
\begin{enumerate}
[label=(\arabic*),itemsep=0pt,topsep=0pt,parsep=0pt]
    \item DV-\yaqi{first} type: DV, CV, DV, CV, ...;
    \item CV-\yaqi{first} type: CV, DV, CV, DV, ....
\end{enumerate}

We define a class of swapping protocols for a genuinely hybrid series QN designed to perform $N$ sequential swapping operations. Each protocol is uniquely characterized by the \emph{swapping order} $s_N= (j_1, j_2, \ldots, j_N)$, specifying the sequence of intermediate nodes $R_{j_1}, R_{j_2}, \ldots, R_{j_N}$ at which entanglement swapping is performed.
Specifically, the selection rule for the $j$-th swapping is determined by the types of the two input states:
\begin{enumerate}
[label=(\arabic*),itemsep=0pt,topsep=0pt,parsep=0pt]
   \item If the two states are of different types, the majorization-based swapping operation is applied. 
   \item If both states are two-qubit states, the DV-based swapping operation \yaqi{in DET (see Table~\ref{table_det})} is used. 
\end{enumerate}
Note that in the swapping process, the presence of DV links fundamentally constrains the effective dimensionality of the distributed entanglement, such that the resulting entangled state is necessarily of DV type.
The system never encounters a configuration where two CV states are connected in series. Consequently, there is no need to invoke a CV-specific swapping operation.

\begin{figure}
    \centering
    \hspace{-15pt}
    \includegraphics[height=125pt]{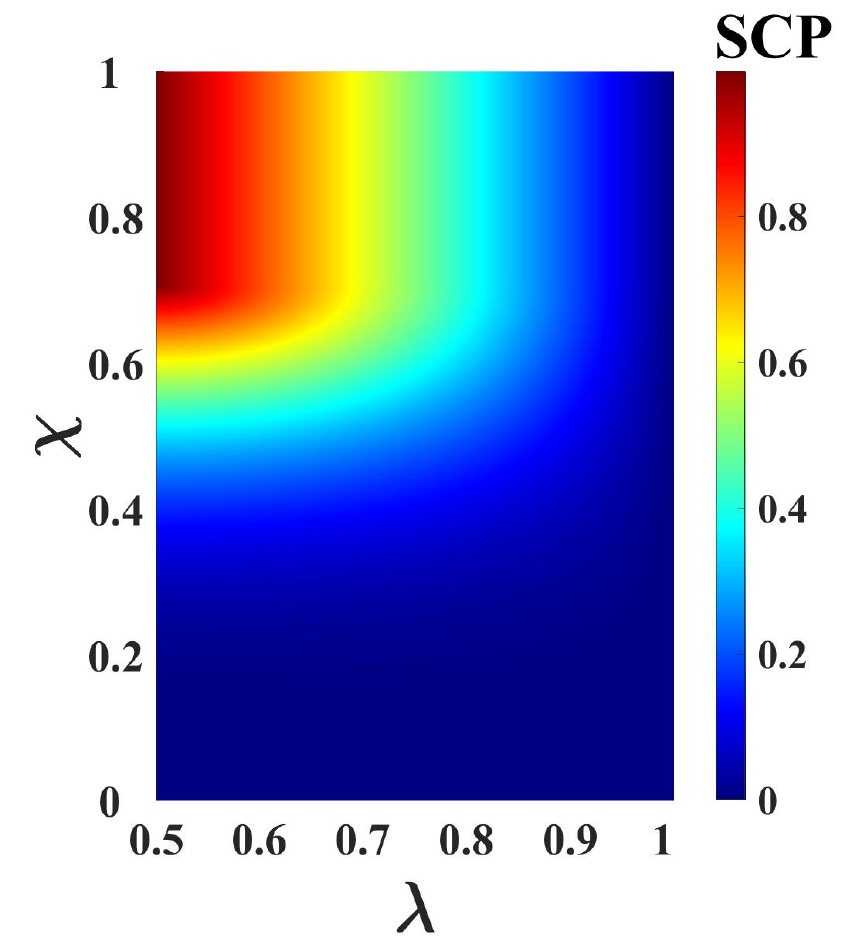}\label{fig-Hybrid_N2}
    \caption{\yaqi{Total SCP achieved by hybrid swapping protocol on a four-node series QN. The topology $S-R_1-R_2-T$ consists of three states, $|\psi^{{\chi}}\rangle_{SR_1} \otimes |\psi_{\lambda}\rangle_{R_1R_2} \otimes |\psi^{{\chi}}\rangle_{R_2T}$.}}
    \label{fig-hybrid_swapping}
\end{figure}

{\color{black}{
Since both operations deterministically produce pure two-qubit states, the resulting entanglement can be characterized by concurrence.
Specifically, each CV resource $|\psi^{\chi_{n_1}}\rangle$ is converted into a two-qubit state with Schmidt coefficient $\eta_{n_1}=\max\{1/2,1-\chi_{n_1}^2\}$ and concurrence $c_{n_1}=2\sqrt{\eta_{n_1}(1-\eta_{n_1})}$, while each DV resource $|\psi_{\lambda_{n_2}}\rangle$ has concurrence $c_{n_2}'=2\sqrt{\lambda_{n_2}(1-\lambda_{n_2})}$. By the DET series rule in Table~\ref{table_det}, the final concurrence is
\begin{equation*}
c
=\prod_{{n_1}=1}^{N_1}c_{n_1}
\prod_{{n_2}=1}^{N_2}c_{n_2}',
\end{equation*}
where $N_1$ and $N_2$ are the numbers of CV and DV links, respectively, satisfying $N_1+N_2=N+1$.

Since $c$ is independent of the swapping order $s_N$, all swapping orders yield the same Schmidt coefficients and the same end-to-end SCP,
\begin{equation}\label{eq-scalable-hybrid-scp}
p_{\mathrm{SCP}}^{\mathrm{seri}}
=1-\sqrt{1-c^2}.
\end{equation}
We collectively refer to these equivalent protocols as the \emph{hybrid swapping protocol}, whose SCP can be evaluated directly from the elementary resource parameters using a number of operations that scales linearly with $N$.}} 
Consider a one-dimensional hybrid QN of CV-first type with $N = 2$ and a state distribution of the form $|\psi^\chi\rangle_{S R_1} \otimes |\psi_\lambda\rangle_{R_1 R_2}\otimes |\psi^\chi\rangle_{R_2 T}$. In Fig.~\ref{fig-hybrid_swapping}, we \yaqi{show the total SCP achieved by} the hybrid swapping protocol.


\section{Genuine Hybrid Concentration}\label{sec-HybridConcentration}

We now develop a majorization-based hybrid concentration protocol for a genuinely hybrid parallel QN, obtained after the \yaqi{homogeneous-first reduction} (Sec.~\ref{sec-hybridQN}).
First, consider a reduced network that consists of exactly two parallel links connecting $S$ and $T$: one DV link described by the state $|\psi_{\lambda}\rangle$ $(\lambda \geq 1/2)$ and one CV link described by the state $|\psi^{{\chi}}\rangle$. 
Based on the majorization theory for state conversions~\cite{Nielsen1999,MRN}, a two-qubit state $|\psi_{\zeta}\rangle$ characterized by $\zeta=\max\left\{{1}/{2},\lambda(1-\chi^2)\right\}$ can be generated (with its entanglement concentrated) from the two source states via LOCC. 
This concentration operation yields a total SCP 
\begin{eqnarray}\label{eq-mu}
    p_{\text{SCP}}=P_{\text{SCP}}(|\psi_{\zeta}\rangle).
\end{eqnarray}
\yaqi{For $\lambda>1/2$ and
$2\sqrt{\zeta(1-\zeta)}> 2\chi/(1-\chi)$, the output has strictly larger entanglement than both inputs.}
This finding demonstrates that DV states can achieve entanglement concentration through synergistic interactions with CV states, and vice versa.

If $\lambda$ and $\chi$ satisfy
$$2\sqrt{\zeta(1-\zeta)}\leq {\frac{2\chi}{1-\chi}},$$
the conversion yields a state $|\psi_{\zeta}\rangle$ with \yaqi{no more} entanglement than $|\psi^{{\chi}}\rangle$. However, converting to the DV state $|\psi_\zeta\rangle$ is still beneficial: it provides greater entanglement than $|\psi_{\lambda}\rangle$ and higher noise resilience for long-distance transmission owing to the inherent stability of DV states. Hence, in this process, $|\psi^{{\chi}}\rangle$ acts as an auxiliary source to boost the entanglement of $|\psi_{\lambda}\rangle$. Note that for $\chi\geq 1/\sqrt{2}$, $|\psi^{{\chi}}\rangle$ can be deterministically transformed into the singlet $|\psi_{1/2}\rangle$ by LOCC, implying that the contribution of $|\psi_{\lambda}\rangle$ can be neglected.

\yaqi{\emph{Scalability.---}In a parallel configuration, an arbitrary number of links connect the same pair of nodes, $S$ and $T$, with each link defined on fixed local Hilbert spaces.
Since the ordering of these states is irrelevant, the elementary concentration operation extends directly to larger parallel QNs.}

\section{Hybrid Entanglement Distribution}\label{sec-DidstributionScheme}

{Based on the hybrid entanglement swapping and hybrid entanglement concentration protocols developed in Secs.~\ref{sec-HybridSwapping} and~\ref{sec-HybridConcentration}, we now assemble a unified entanglement distribution scheme for genuinely hybrid series-parallel QNs. To evaluate its performance, we compare the proposed scheme with a benchmark scheme, both on series-parallel networks and on the honeycomb lattice which serves as a paradigmatic example of non-series-parallel topology.}

\subsection{Scheme}

Any series-parallel network can be decomposed into nested combinations of series and parallel sub-networks, implying that the entanglement distribution in arbitrary hybrid series-parallel QNs can be achieved by iteratively applying the hybrid protocols for series and parallel configurations. 
Consequently, the hybrid swapping protocol (for series topologies) and the hybrid concentration protocol (for parallel topologies) together form a unified and scalable framework---termed the \emph{hybrid entanglement distribution} scheme---for efficient entanglement distribution across series-parallel networks.

To demonstrate its implementation, we consider a hybrid QN with a series-then-parallel topology [Fig.~\ref{fig-s-t-p}] as a representative example in the following analysis.

\emph{Example.---}Consider a simplified hybrid QN with a series-then-parallel topology [Fig.~\ref{fig-s-t-p}] between nodes $S$ and $T$, composed of multiple series sub-networks $\mathcal{G}_{N_1}, \mathcal{G}_{N_2}, \dots, \mathcal{G}_{N_K}$ connected in parallel. 
Applying the hybrid swapping protocol, each sub-network $\mathcal{G}_{N_k}$ generates \yaqi{a state} $|\psi_{\lambda_k}\rangle$ between $S$ and $T$. \yaqi{Subsequent deterministic entanglement concentration of these $K$ states produces the final state $|\psi_\lambda\rangle$, where $\lambda=\max\{1/2,\prod_{k}\lambda_k\}$.
This yields a total SCP 
$$p_{\text{SCP}}=P_{\rm SCP}(|\psi_\lambda\rangle)=\min\{1,2-2\prod_{k}\lambda_k\}.$$
}

\subsection{Efficiency analysis}

{To assess the efficiency of the proposed hybrid entanglement distribution scheme, it is necessary to introduce an appropriate benchmark protocol for comparison. To this end, we consider a benchmark scheme based on singlet conversion, which provides a topology-agnostic baseline widely adopted in entanglement percolation studies~\cite{Percolation_CEP2007,XJS2021,G2_conversion2025}.}

{\emph{Pairwise (point-to-point) singlet conversion (PSC) scheme.---}The scheme involves two steps:
\begin{enumerate}
[label=(\arabic*),itemsep=0pt,topsep=0pt,parsep=0pt]
    \item Independently convert each adjacent, point-to-point entangled pair into a singlet (maximally entangled state) via LOCC, with a success probability defined by the SCP of the state.
    \item Upon successful conversion of all states along a path from $S$ to $T$, perform the Bell-state measurement at each intermediate node to deterministically create an end-to-end singlet between $S$ and $T$.
\end{enumerate}
Through this procedure, the hybrid QN becomes a random graph~\cite{random_graph1959}, where each link exists with a probability equal to the SCP of its corresponding entangled state. The end-to-end connection probability between $S$ \yaqi{and} $T$ in this graph is equivalent to the probability of successful end-to-end singlet generation---that is, the total SCP. The corresponding series and parallel rules for the scheme \yaqi{are} summarized in Table~\ref{table_LSC}.}

    \begin{table}[]
    \caption{Pairwise (point-to-point) singlet conversion.}
    \label{table_LSC}
    \begin{tabular}{c|c|c}
		\hline\hline
		    & Series rule & Parallel rule \\
          \hline
          End-to-end SCP & $p=p_1p_2\cdots$ & $1-p=(1-p_1)(1-p_2)\cdots$\\
		\hline\hline
	\end{tabular}
\end{table}

{In the following, we compare the efficiency of the proposed hybrid entanglement distribution scheme with that of the PSC benchmark on the series-parallel network and the honeycomb lattice. These two cases allow us to assess the performance of the two schemes both within and beyond the class of series-parallel networks.
\yaqi{We also restrict our comparison to genuinely hybrid QNs by appealing to the homogeneous-first reduction.}}


\subsubsection{Series-parallel networks}\label{sec-hybrid_protocol}

The (two-terminal) series-parallel networks are two-terminal structures generated recursively from single-component networks by successive series and parallel compositions~\cite{Duffin1965,SP1942,SP1972}.
Starting from all one-component structures, increasingly complex networks are obtained by repeatedly combining previously constructed, component-distinct sub-networks in series or in parallel a finite number of times.
This constructive definition yields a well-defined class of two-terminal networks, which underlies the hierarchical organization exploited throughout this section.

We now exploit this intrinsic decomposability of series-parallel networks to analyze the genuinely hybrid QN in terms of series and parallel components.

\emph{Series components.---}\yaqi{Consider} a genuinely hybrid series QN composed of $N_1$ TMSVSs $|\psi^{{\chi_{n_1}}}\rangle$ ($n_1=1,2,...,N_1$) and $N_2$ two-qubit states $|\psi_{\lambda_{n_2}}\rangle$ ($n_2=1,2,...,N_2$) where $N_1+N_2=N{+1}$ and $|N_1-N_2|\leq1$.
We now demonstrate the advantage of the hybrid swapping protocol over the PSC protocol.

For any $N\geq1$,
the \yaqi{hybrid} swapping protocol deterministically generates a two-qubit state $|\psi_{\lambda}\rangle$ satisfying
\begin{eqnarray*}
   \lambda=\frac{1}{2}\left(1+\sqrt{1-\prod_{n_1=1}^{N_1}c_{n_1}{^2}\prod_{n_2=1}^{N_2}(c'_{n_2}){^2}}\right),
\end{eqnarray*}
where $c_{n_1}=2\sqrt{\eta_{{n_1}}(1-\eta_{{n_1}})}$ {with $\eta_{{n_1}}=\max\left\{{1}/{2}, 1-{\chi^2_{n_1}}\right\}$} and $c'_{n_2}=2\sqrt{\lambda_{n_2}(1-\lambda_{n_2})}$.
The corresponding total SCP  is 
\begin{eqnarray*}
    {p_{\text{hybrid}}^{\yaqi{\rm seri}}}=P_{\text{SCP}}(|\psi_{\lambda}\rangle)=1-\sqrt{1-\prod_{n_1=1}^{N_1}c_{n_1}{^2}\prod_{n_2=1}^{N_2}(c'_{n_2}){^2}}.\nonumber
\end{eqnarray*}

Under the PSC protocol (Table~\ref{table_LSC}), the total SCP $p^{\text{seri}}_{\text{PSC}}$ is the product of the individual SCPs of all initial states,  
\begin{eqnarray*}
    p^{\text{seri}}_{\text{PSC}}&=&\prod_{n_1=1}^{N_1}P_{\text{SCP}}(|\psi^{{\chi_{n_1}}}\rangle)\prod_{n_2=1}^{N_2}P_{\text{SCP}}(|\psi_{\lambda_{n_2}}\rangle)\nonumber\\&=&
    2^{{N+1}}\prod_{n_1=1}^{N_1}(1-\eta_{{n_1}})\prod_{n_2=1}^{N_2}(1-\lambda_{{n_2}}).
\end{eqnarray*}
We find that the inequality
\begin{eqnarray}\label{eq-advantage_seri}
    p^{\text{seri}}_{\text{PSC}}\leq {p_{\text{hybrid}}^{\yaqi{\rm seri}}}
\end{eqnarray}
\yaqi{always} holds {(Appendix~\ref{app-seri_compare})}, demonstrating that the \yaqi{hybrid} swapping protocol  \yaqi{performs at least as well as} the PSC in terms of total SCP. 

\emph{Parallel topology.---}\yaqi{Consider} a genuinely hybrid parallel QN where two nodes $S$ and $T$ share a two-qubit state $|\psi_{\lambda}\rangle$ and a TMSVS $|\psi^{{\chi}}\rangle$. Under the PSC protocol, the total SCP for the parallel configuration is given by
\begin{eqnarray*}
    p_{{\text{PSC}}}^{\text{para}}
    =1-[1-P_{\text{SCP}}(|\psi_{\lambda}\rangle)][1-P_{\text{SCP}}(|\psi^{{\chi}}\rangle)].
\end{eqnarray*}
It follows that
\begin{eqnarray}\label{eq-advantage_para}
    p_{{\text{PSC}}}^{\text{para}}\leq {p_{\text{hybrid}}^{\rm para}}
\end{eqnarray}
where $p_{\text{hybrid}}^{\rm para}=P_{\text{SCP}}(|\psi_{\zeta}\rangle)$ [Eq.~\eqref{eq-mu}] is the total SCP induced by the hybrid concentration protocol, indicating that the majorization-based LOCC conversion enhances the SCP in the QN.
This enhancement originates from the fact that majorization-based LOCC conversion enables parallel entangled resources to be processed jointly and conditionally. 
Through such collective processing, weak entanglement initially distributed over multiple resources can be concentrated into a single, stronger entangled state. 
By contrast, the PSC protocol treats each entangled pair independently, preventing any integration across parallel resources.

The demonstrated advantage of the hybrid distribution scheme in both series [Eq.~\eqref{eq-advantage_seri}] and parallel networks [Eq.~\eqref{eq-advantage_para}] establishes its high efficiency across all {genuinely hybrid {series and parallel} QNs}. These results provide a basis for applying the proposed resource-level framework to more general hybrid series-parallel QNs and suggest its potential for reducing resource requirements as the network scales.

\subsubsection{Honeycomb lattice}\label{sec-honeycomb}

\begin{figure}[]
    \centering
    \subfigure[]{
    \includegraphics[height=80pt]{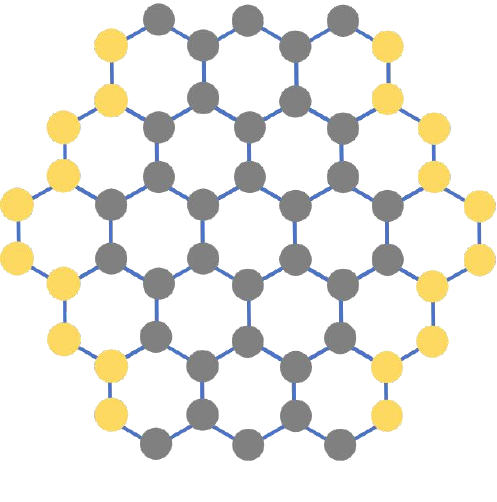}\label{fig-honeycomb1}
    }\hspace{-5pt}
    \subfigure[]{
    \includegraphics[height=80pt]{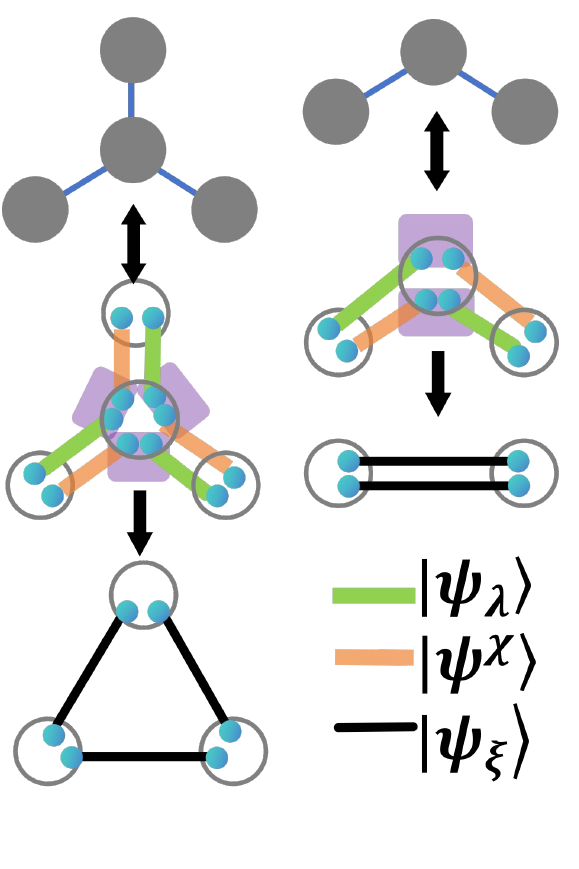}\label{fig-distribution}
    }\hspace{-5pt}
    \subfigure[]{
    \includegraphics[height=80pt]{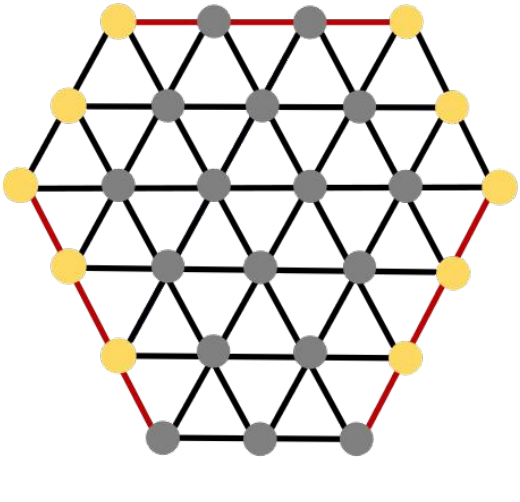}\label{fig-trangular_hybrid}
    }

    \subfigure[]{\hspace{-12pt}\includegraphics[height=100pt]{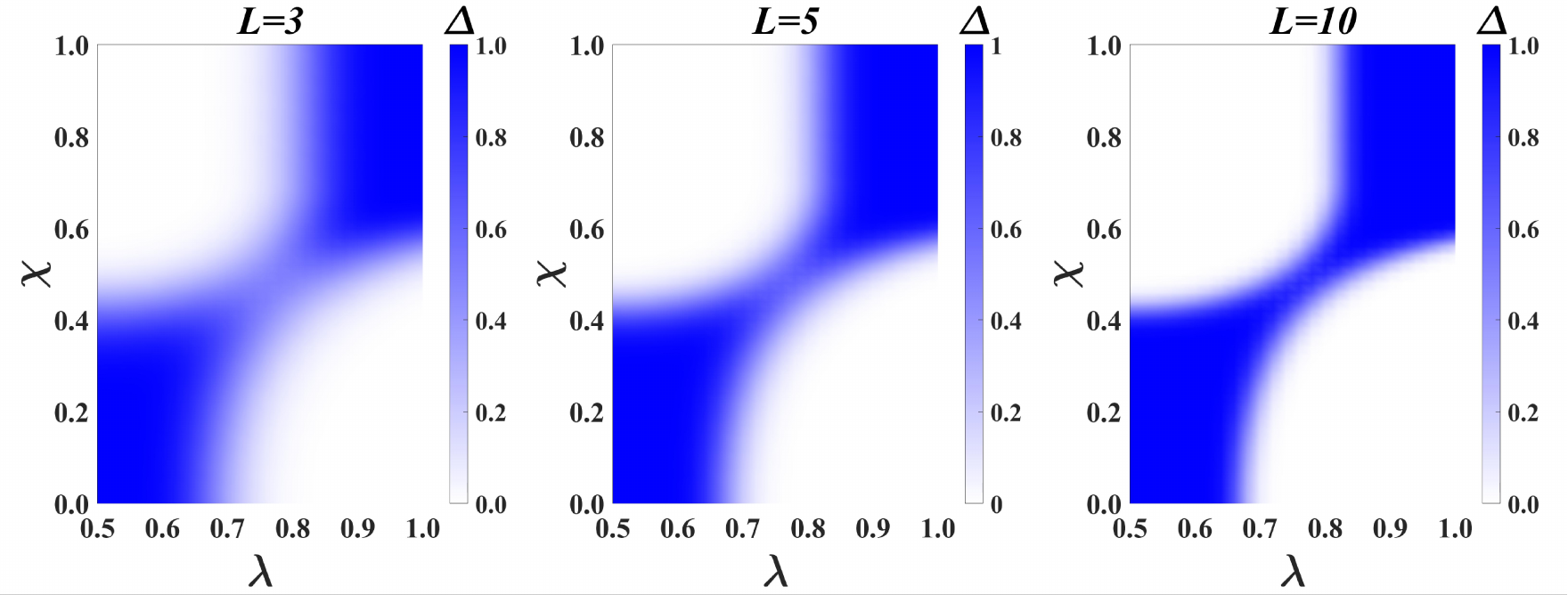}\label{fig-Delta_L}
    }\vspace{-2mm}
    \caption{{Entanglement distribution operations on hybrid double-link honeycomb lattice.}
    \subref{fig-honeycomb1}~We consider a finite-size double-link honeycomb lattice with $L=3$ layers extended outward from a central hexagon. \subref{fig-distribution}~Each blue segment is a shorthand for two states in parallel---$|\psi_\lambda\rangle$ (green link) and $|\psi^{{\chi}}\rangle$ (orange link). Therefore, each trifurcation (or bifurcation when on the boundary) can be decomposed into 3 (or 2) pairs of subsystems  (purple region). {The hybrid swapping pre-processing is then implemented by performing entanglement swapping on every such pair of subsystems, after which the lattice transforms into a triangular lattice.} \subref{fig-trangular_hybrid}~In the triangular lattice, each black link represents a state $\vert\psi_{\yaqi{\xi}}\rangle$, and each red segment consists of two parallel black links.
    Let $\Delta = P_{\text{SC}}^{\mathcal{G}_{\yaqi{1}}} - P_{\text{SC}}^{\mathcal{G}_{\yaqi{2}}}$ denote the difference in total SCP between the PSC protocol \yaqi{without} ($P_{\text{SC}}^{\mathcal{G}_{\yaqi{1}}}$) and \yaqi{with} ($P_{\text{SC}}^{\mathcal{G}_{\yaqi{2}}}$) hybrid swapping pre-processing. \subref{fig-Delta_L}~For finite honeycomb lattices ($L=3,\yaqi{5,10}$), $\Delta$ is \yaqi{always positive}, {indicating that the hybrid pre-processing does not enhance efficiency.}}
    \label{fig-Honeycomb}
\end{figure}

{To study the role of hybrid quantum operations in {genuinely hybrid non–series-parallel QNs}, we consider a honeycomb lattice [Fig.~\ref{fig-honeycomb1}], a prototypical geometry that cannot be reduced to series-parallel form. The two opposite boundaries of the lattice (yellow nodes) are connected through multiple paths composed of hybrid entangled links. Each link consists of a two-qubit state $|\psi_{\lambda}\rangle$ (green) and a two-mode squeezed vacuum state $|\psi^{{\chi}}\rangle$ (orange), as shown in Fig.~\ref{fig-distribution}. We compare two mapping strategies that reduce the hybrid lattice to an effective classical random graph, enabling a quantitative analysis of end-to-end entanglement connectivity.}

(1) \emph{PSC protocol.---}{Under the PSC protocol, each hybrid link $|\varphi\rangle=|\psi_{\lambda}\rangle\otimes|\psi^{{\chi}}\rangle$ is independently converted into a singlet with probability}
\yaqi{\begin{equation*}
    q_{\text{PSC}}=1-[1-P_{\text{SCP}}(|\psi_{\lambda}\rangle)][1-P_{\text{SCP}}(|\psi^{{\chi}}\rangle)].
\end{equation*}}
As a result, the original entanglement distribution problem in honeycomb lattice is mapped onto a classical bond-percolation problem, where each link of the lattice is retained with probability $q_{\text{PSC}}$ and removed otherwise. Entanglement distribution across the lattice is therefore determined by the connectivity properties of this randomly diluted honeycomb lattice denoted by $\mathcal{G}_1$.

(2) \emph{Hybrid swapping protocol.---}{In the hybrid swapping approach, entanglement distribution on the honeycomb lattice is carried out by locally applying hybrid entanglement swapping operations before PSC. 
At each trifurcating node of one selected sublattice of the honeycomb lattice, the six incoming subsystems are grouped into three hybrid pairs [purple blocks in Fig.~\ref{fig-distribution}]. Each hybrid pair shares one two-qubit state $|\psi_\lambda\rangle$ and one TMSVS $|\psi^{\chi}\rangle$ with two neighboring nodes. As shown in Sec.~\ref{sec-HybridSwapping}, applying the hybrid swapping protocol to each such pair generates a two-qubit state {in the form of $|\psi_{\yaqi{\xi}}\rangle=\sqrt{\yaqi{\xi}}|01\rangle+\sqrt{1-{\yaqi{\xi}}}|10\rangle$} [black link in Fig.~\ref{fig-distribution}]. The corresponding total SCP of this effective link is therefore
\begin{eqnarray*}
    q^{\text{hybrid}}= P_{\text{SCP}}(|\psi_{\yaqi{\xi}}\rangle)\yaqi{=2(1-\xi)}.
\end{eqnarray*}
At the remaining bifurcating nodes that are not involved in the first swapping step, the four subsystems are similarly grouped into two hybrid pairs [Fig.~\ref{fig-distribution}]. Applying the hybrid swapping protocol to each pair yields two independent black links in parallel. These two links are equivalently represented as a single effective link [red link in Fig.~\ref{fig-trangular_hybrid}] with connectivity probability $q=1-(1-q^{\text{hybrid}})^2$.
Once all links along a path connecting the two opposite boundaries (yellow nodes) are successfully converted into singlets, BSMs performed at the intermediate nodes establish an end-to-end singlet between the two boundaries, thereby confirming successful entanglement distribution.
Overall, this procedure maps the original hybrid QN to a random network $\mathcal{G}_2$ defined on a triangular lattice with two boundaries, as shown in Fig.~\ref{fig-trangular_hybrid}. In this effective network, each black link exists independently with probability $q^{\text{hybrid}}$, while each red link exists with probability $q$. Consequently, the entanglement distribution problem reduces to determining the connectivity between the two boundaries of $\mathcal{G}_2$.}

Let $P_{\text{SC}}^{\mathcal{G}_1}$ and $P_{\text{SC}}^{\mathcal{G}_2}$, respectively, denote the total end-to-end SCP of the random networks $\mathcal{G}_1$ and $\mathcal{G}_2$, where each is defined as the probability that at least one path connects the two boundaries. The performance difference between the hybrid swapping protocol and the PSC protocol is quantified as
\begin{eqnarray*}
    \Delta=P_{\text{SC}}^{\mathcal{G}_{\yaqi{1}}}-P_{\text{SC}}^{\mathcal{G}_{\yaqi{2}}}.
\end{eqnarray*}
For a finite-size honeycomb lattice of size $L$ (number of layers extended from a central hexagon), Monte Carlo simulations reveal that the inequality $\Delta\yaqi{\geq}0$ holds {true} [Fig.~\ref{fig-Delta_L}]. This suggests that the hybrid swapping protocol offers no practical advantage over the PSC protocol in such finite-size networks. 
In the limit $L\to\infty$, \yaqi{the percolation-threshold analysis in Appendix~\ref{app-honeycomb} shows that the supercritical region ($P_{\text{SC}}^{\mathcal{G}_{\yaqi{2}}}>0$) of the hybrid-swapping preprocessing protocol is contained within that ($P_{\text{SC}}^{\mathcal{G}_{\yaqi{1}}}>0$) of the PSC protocol,} demonstrating that the hybrid swapping does not enhance the entanglement distribution efficiency in honeycomb lattices.
{This limitation is structural {within hybrid entanglement swapping}. Since all scalable hybrid swapping operations collapse CV resources into two-qubit DV states, the effective bond connectivity in the resulting mapped graph cannot exceed that obtained via direct PSC conversion in \yaqi{specific} non-series-parallel geometries.}

{This finding exposes \yaqi{topology-dependent performance within the present reduction framework} in hybrid entanglement distribution. 
While the hybrid swapping protocol is explicitly designed to exploit the recursive structure of series-parallel networks and therefore achieves an advantage over the PSC benchmark in such topologies, its use as a pre-processing step does not improve---and can even degrade---the entanglement distribution efficiency in certain non–series-parallel configurations. 
Consequently, the performance gain enabled by hybrid swapping in series-parallel networks does not generally extend to non–series-parallel architectures.
This behavior indicates that the advantage of hybrid entanglement distribution protocols is intrinsically topology-dependent rather than universal, being closely tied to whether the underlying network admits a recursive series-parallel decomposition.}

{In non-series-parallel networks, this structural compatibility is absent, and alternative swapping strategies---particularly those that generate \yaqi{ensembles of conditional output states} and \yaqi{are difficult to analyze exactly in larger series networks due to the growing ensemble complexity}---may instead yield higher distribution efficiency, as demonstrated in Appendix~\ref{app-honeycomb}.}

\section{Resource estimation}
\label{sec-implementation}

{The analysis developed above assumes ideal pure-state resources and perfect local operations. In particular, we neglect state-preparation errors, channel loss, detector inefficiency, quantum-memory decoherence, and the probabilistic nature of DV entanglement generation. For practical applications, our framework should therefore be interpreted as a resource-level baseline rather than a complete operational model. Its principal advantage is that it separates the network-level transmission rules from the microscopic implementation of each elementary link. Once the entanglement quality and availability of these basic resources are characterized in the lab, they may be plugged directly into the series and parallel reduction procedures developed in the preceding sections.}

{For a CV link described by a TMSVS, the elementary resource is parameterized by $\chi$. For reference, a squeezing level of  $10\,\mathrm{dB}$ corresponds to $\chi\approx0.82$. Recent optical experiments have achieved squeezing beyond this threshold, alongside the network-compatible generation and distribution of CV entanglement~\cite{Vahlbruch2016,Larsen2019,CV_QN2024,CV2025,CV_code_chip2025}. For a DV link described by a two-qubit state, the corresponding resource is parameterized by the Schmidt coefficient $\lambda$. For reference, state-of-the-art heralded remote entanglement experiments typically achieve Bell-state fidelities $\sim 90\%$, which corresponds to an effective Schmidt parameter $\lambda \approx 0.80$.
In practice, $\lambda$ is inferred from state characterization after a successful link is established, and it often depends on factors such as source brightness, heralding probability, channel transmission, detector efficiency, and memory survival rates~\cite{Awschalom2021}.}

{Distance-dependent attenuation can also be folded into this resource-level picture. For example, the transmission efficiency of an optical channel of length $L$ is commonly modeled as $\eta(L)=e^{-L/L_{\rm att}}$, where $L_{\rm att}$ denotes the attenuation length~\cite{Pirandola2017,Takeoka2014}.
\yaqi{To illustrate the effect of distance-dependent resource degradation within our framework, we introduce effective parameters $\chi_{\rm eff}(L)$ and $\lambda_{\rm eff}(L)$ for the CV and DV resources, respectively, and substitute them into the network-level formulas:}
\begin{eqnarray}\label{substitutions}
    \chi\rightarrow\chi_{\rm eff}(L),\qquad
\lambda\rightarrow\lambda_{\rm eff}(L).
\end{eqnarray}
\yaqi{These substitutions 
assume an idealized filtering or purification (possibly with auxiliary registers and postselection) that maps the transmitted link back to the same pure-state family used in the network-level reduction, yielding effective parameters $\chi_{\rm eff}(L)$ and $\lambda_{\rm eff}(L)$. If the filtering or postselection is probabilistic, the effective parameters $\chi_{\rm eff}(L)$ and $\lambda_{\rm eff}(L)$ represent conditional effective parameters for a successful postselection event. 
For non-unitary loss, this is an approximation rather than an exact mixed-state treatment.
If the output states after the loss channel cannot be recovered to the form of the initial states, or if the performance significantly depends on additional variables (such as generation rate or memory lifetime), a single effective parameter is insufficient, and a mixed-state or rate-dependent model is required.}


These single, effective parameters provide a scalable, hardware-aware workflow for resource estimation. First, each elementary link is characterized by its effective entanglement parameter and availability probability. Next, homogeneous subgraphs are reduced using the resource-specific rules from Sec.~\ref{sec-hybridQN}. Finally, hybrid swapping and concentration protocols are applied across the remaining DV--CV interfaces. The resulting end-to-end SCP can be compared with a prescribed performance target, allowing us to estimate the required elementary-link quality, number of parallel resources, or degree of multiplexing without modifying the underlying network-level framework \yaqi{(with examples given in Appendix~\ref{sec-AP-examples})}.}

{The effective-resource description clarifies the complementary roles of CV and DV links in a practical hybrid QN. While CV entanglement can typically be generated at high rates, its effective entanglement quality drops rapidly in the presence of optical loss. In contrast, DV entanglement is usually generated stochastically at lower rates, but successful heralding events can preserve high-quality entanglement that is better suited for long-distance transmission and quantum storage. Therefore, a practical network design will likely rely on abundant CV links with small squeezing parameters $r$, bridged by fewer DV links with high entanglement (i.e., low Schmidt coefficients $\lambda$). Our framework provides a common language to compare these disparate systems and identify which resource type limits the end-to-end performance. If hardware noise produces states that cannot be accurately captured by a single effective parameter, a more explicit mixed-state and rate-dependent model will be necessary. Even then, the present model serves as a tractable and intuitive baseline for those more detailed analyses.}

\section{Conclusion and discussion}\label{sec-conclusion}

In this work, we established a theoretical framework for entanglement distribution in hybrid pure-state quantum networks (QNs) that incorporate both DV and CV entangled sources. By modeling architectures with specific series-parallel topologies, we demonstrate that targeted hybrid quantum operations can enhance distribution efficiency. This result showcases a definitive advantage within the class of genuinely hybrid series-parallel QNs, where the integration of diverse encodings outperforms traditional single-encoding benchmarks.

Despite these advancements, several challenges remain for future research:
\begin{enumerate}
\item While our proposed hybrid protocol offers advantages over traditional benchmarks, it is likely not globally optimal. Determining whether more efficient (but potentially not scalable) hybrid protocols exist remains a valuable direction for investigation.
\item For complex series-parallel networks, multiple different decompositions are often possible. Ensuring the optimality of a specific nested decomposition into series and parallel rules poses significant theoretical and computational challenges.
\item Developing hybrid protocols for general, non-series-parallel network topologies (such as lattices or random graphs) remains largely unexplored. Our findings on the honeycomb lattice suggest that these geometries may require fundamentally different strategies to surpass point-to-point benchmarks.
\item 
{Incorporating realistic physical imperfections, including channel attenuation, imperfect state preparation, detector inefficiency, and quantum-memory decoherence, into the hybrid entanglement distribution framework is an important step toward quantitative performance analysis for practical QNs.}
\end{enumerate}

{Beyond its theoretical significance, the present framework provides a resource-level description of hybrid QNs that is independent of the underlying hardware platform. Under certain conditions, realistic experimental imperfections can be incorporated through effective resource parameters, allowing the analytical results developed here to be applied to practical network architectures once the elementary entangled resources have been experimentally characterized. In this sense, the framework serves not only as a theoretical model for hybrid entanglement distribution but also as a scalable performance-prediction tool for heterogeneous QNs. Future work combining the present resource-based description with hardware-specific models, including realistic loss, quantum-memory decoherence, multiplexing techniques, and quantum repeater protocols, may provide quantitative guidance for the design and optimization of large-scale hybrid quantum internet architectures.}

\section*{ACKNOWLEDGMENT}

Y.Z. (Zhao), K.H., and J.H. were supported by the National Natural Science Foundation of China under Grants No.~12271394 and No.~12071336. 
N.L.P. was supported by the JSPS KAKENHI Grants No.~24K07485.

\newpage
\clearpage

\section*{APPENDIX}

\renewcommand{\thesection}{Appendix}
\numberwithin{table}{section}
\renewcommand{\thetable}{\thesection\arabic{table}}
\appendix

{\color{black}
\section{{An alternative homodyne-based hybrid swapping protocol}}\label{app-homodyne}

An alternative entanglement swapping between DV and CV states is achievable for any non-zero squeezing of the CV resource (i.e.,~{$\chi=\tanh r > 0$}), as established in Refs.~\cite{RT1999,Takeda2013_1,Takeda2013_2,SMPA2015}. In this setup, the two-qubit state $|\psi_{\lambda}\rangle$ is prepared by passing a single photon through a beam splitter with reflectivity $1-\lambda$, while the TMSVS $|\psi^{{\chi}}\rangle$ is generated from the vacuum by a two-mode squeezing operator with squeezing parameter $r=\text{arctanh}\, \chi>0$. Since both DV and CV states can be represented in the Fock basis, performing homodyne measurement with a feedforward gain $g=\chi$ maps $|\psi_{\lambda}\rangle_{SR}\otimes|\psi^{{\chi}}\rangle_{RT}$ deterministically onto a mixed two-qubit state
\begin{equation}\label{CV-swap}
  \varrho_{ST}(\lambda,{\chi})=\alpha|\psi_{\beta}\rangle_{ST}\langle\psi_{\beta}|+(1-\alpha)|00\rangle_{ST}\langle00|
\end{equation}
with $\alpha=1-\lambda(1-\chi^2)$ and $\beta=\lambda\chi^2/\alpha$, respectively~\cite{SMPA2015}. 
Equation~\eqref{CV-swap} shows that the homodyne-based swapping operation deterministically produces a mixed two-qubit state, consisting of a pure entangled component $|\psi_{\beta}\rangle_{ST}$ with weight $\alpha$ and a vacuum component $|00\rangle_{ST}$ with weight $1-\alpha$. The SCP of the entangled component is $P_{\mathrm{SCP}}(|\psi_{\beta}\rangle_{ST})=2\min\{\beta,1-\beta\}$. Accordingly, we introduce the weighted SCP of the entangled component as
\begin{eqnarray}\label{eq-homo_SCP}
p^{\mathrm{homo}}
&=&
\alpha P_{\mathrm{SCP}}(|\psi_{\beta}\rangle_{ST})\nonumber\\
&=&2\min\{\lambda\chi^2,1-\lambda\}.
\end{eqnarray}
Note that this characterizes the weighted SCP of the entangled component rather than the operational SCP of the unheralded mixed output $\varrho_{ST}$.}


\yaqi{This protocol provides another example that entanglement can be transferred across a DV--CV interface into a two-qubit system.
However, since the resulting state is generally mixed, its incorporation into the pure-state entanglement distribution framework considered here would require an additional purification or suitable heralded state-preparation procedure.}

\section{Comparison of the probabilities ${p_{\text{hybrid}}^{\yaqi{\rm seri}}}$ and $p^{\text{seri}}_{\text{PSC}}$}\label{app-seri_compare}

{We now prove the inequality 
\begin{eqnarray}
    {p_{\text{hybrid}}^{\yaqi{\rm seri}}}\geq {p^{\text{seri}}_{\text{{PSC}}}}.
\end{eqnarray}
It is sufficient to show that
\begin{eqnarray}\label{eq-app_comparison}
    \left(1-{p^{\text{seri}}_{\text{{PSC}}}}\right)^2-\left(1-{p_{\text{hybrid}}^{\yaqi{\rm seri}}}\right)^2 \geq 0,
\end{eqnarray}
where
\begin{eqnarray*}
    \left(1-{p^{\text{seri}}_{\text{{PSC}}}}\right)^2-\left(1-{p_{\text{hybrid}}^{\yaqi{\rm seri}}}\right)^2=f^{\text{seri}}h^{\text{seri}},
\end{eqnarray*}
with
\begin{eqnarray*}
    f^{\text{seri}}&=&\prod_{n_1}2(1-\eta_{n_1})\prod_{n_2}2(1-\lambda_{n_2}),\\
    h^{\text{seri}}&=&f^{\text{seri}}+\prod_{n_1}2\eta_{n_1}\prod_{n_2}2\lambda_{n_2}-2.
\end{eqnarray*}}

{For any $n_1=1,2,\ldots,N_1$,
\begin{eqnarray*}
    \frac{\partial h^{\text{seri}}}{\partial \eta_{n_1}}&=&2^{N+1}\Big[-\prod_{n_1'\neq n_1}(1-\eta_{n_1'})\prod_{n_2}(1-\lambda_{n_2})\\
    &&\ \ \ \ \ \ \ +\prod_{n_1'\neq n_1}\eta_{n_1'}\prod_{n_2}\lambda_{n_2}\Big]\\
    &\geq& 0,
\end{eqnarray*}
where the inequality follows since $\eta_{n_1'},\lambda_{n_2}\geq 1/2$. Likewise,
\begin{eqnarray*}
    \frac{\partial h^{\text{seri}}}{\partial \lambda_{n_2}}\geq 0
\end{eqnarray*}
for every $n_2=1,2,\ldots,N_2$.}

{Therefore, $h^{\text{seri}}$ is monotonically increasing with respect to each variable and attains its minimum value,
\begin{eqnarray*}
    h^{\text{seri}}=0,
\end{eqnarray*}
at $\eta_{n_1}=\lambda_{n_2}=1/2$ for all $n_1$ and $n_2$. This implies that  $h^{\text{seri}}\geq0$, so Eq.~\eqref{eq-app_comparison} immediately follows, which completes the proof.}

\section{Entanglement distribution efficiency in the infinite honeycomb lattice}\label{app-honeycomb}

{Prior work has employed the framework of entanglement percolation to investigate the efficiency of distributing entanglement between two distant boundaries across large-scale QNs with non-series-parallel topologies~\cite{Percolation_CEP2007,percolation2008,percolation2009,Percolation_mixed2010,percolation_2013,XJS2021}. In a QN, under a scalable entanglement distribution scheme, initial entanglement resources are converted into end-to-end ``sponge-crossing'' entanglement~\cite{XJS2021,NegPT}, which serves as a quantification of distribution efficiency. For an infinitely extended QN composed of entangled states with identical entanglement, percolation theory provides an effective analytical tool, with a key metric being the entanglement threshold---the minimum initial entanglement required for non-zero sponge-crossing entanglement~\cite{Percolation_CEP2007}. A lower threshold implies that the network can operate efficiently with weaker initial entanglement, thereby reducing resource waste and improving overall distribution efficiency.}

{To investigate the impact of hybrid quantum operations on entanglement distribution in large-scale \yaqi{networks}, we focus on the entanglement percolation model in an infinite-size honeycomb lattice. Our goal is to evaluate whether this hybrid protocol can lower the percolation threshold.}

We now prove that the hybrid swapping protocol does not \yaqi{provide a uniform improvement over PSC} in an infinite honeycomb lattice.

\begin{figure}
    \centering
    \includegraphics[width=250pt]{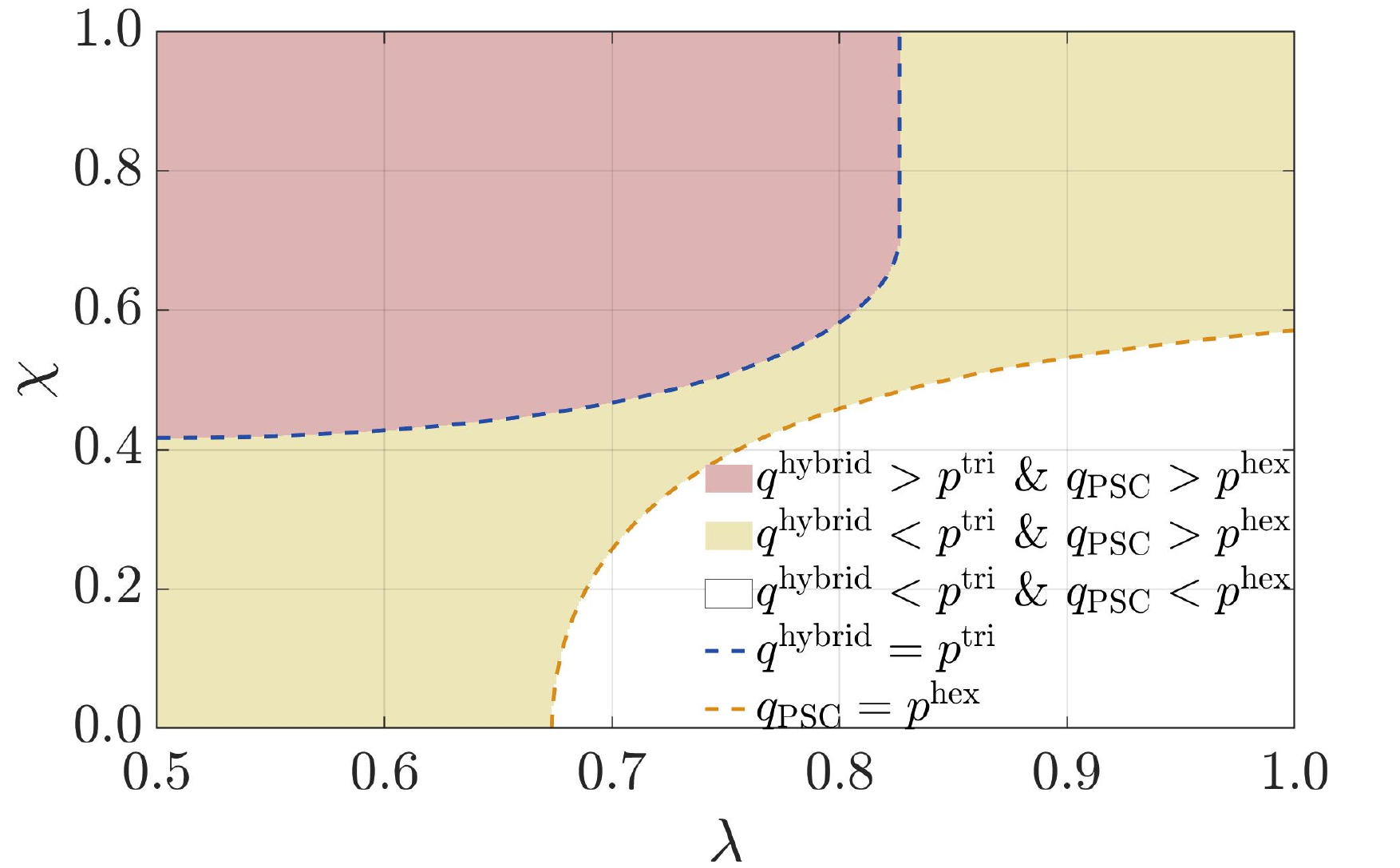}
    \caption{\yaqi{Percolation regions for the PSC protocol without and with hybrid-swapping pre-processing in an infinite honeycomb lattice. For these two cases, let  $P_{\text{SC}}^{\mathcal{G}_{1}}$ and $P_{\text{SC}}^{\mathcal{G}_{2}}$ denote the total SCPs. For the former, the pink and yellow areas jointly constitute the percolation regions corresponding to $P_{\text{SC}}^{\mathcal{G}_{1}}>0$ (i.e., $q_{\rm PSC} >p^{\rm hex}$), with the orange dashed line marking the threshold boundary $q_{\rm PSC}=p^{\rm hex}$. For the latter, the pink areas represent the percolation regions corresponding to $P_{\text{SC}}^{\mathcal{G}_{2}}>0$ (i.e., $q^{\rm hybrid}  > p^{\rm tri}$), with the blue dashed line marking the threshold boundary $q^{\rm hybrid} = p^{\rm tri}$. This region lies within the former’s percolation region.}}
    \label{fig_honeycomb_Linfty}
\end{figure}

\yaqi{In the limit of infinite network size, $L\to\infty$,} the percolation threshold \yaqi{of} $q^{\text{hybrid}}$ \yaqi{is} the known bond percolation threshold
\begin{eqnarray*}
    p^{\text{tri}} = 2\sin{(\pi/18)}
\end{eqnarray*}
for a triangular lattice~\cite{Triangular_honeycomb1964}. 
This threshold, in turn, defines the critical threshold pair $(\lambda_{\text{th}}^{\text{hybrid}},\chi_{\text{th}}^{\text{hybrid}})$ of $(\lambda,\chi)$.
\yaqi{
\begin{eqnarray*}
  &&\lambda_{\text{th}}^{\text{hybrid}}\left(1-\lambda_{\text{th}}^{\text{hybrid}}\right)
  =\frac{1-\left(1-2\sin{\frac{\pi}{18}}\right)^2}{16\eta^{\text{hybrid}}(1-\eta^{\text{hybrid}})},\\
  &&\eta^{\text{hybrid}}=\max\left\{\frac{1}{2},1-\left(\chi_{\text{th}}^{\text{hybrid}}\right)^2\right\}.
\end{eqnarray*}}
\yaqi{The set of all threshold points $(\lambda_{\text{th}}^{\text{hybrid}},\chi_{\text{th}}^{\text{hybrid}})$ forms the boundary indicated by the blue dashed line in Fig.~\ref{fig_honeycomb_Linfty}.}

Under the PSC protocol, 
the corresponding percolation threshold $q_{\text{PSC}}$ is the bond percolation threshold of the honeycomb lattice~\cite{Triangular_honeycomb1964}:
\begin{eqnarray*}
    p^{\text{hex}}=1-2\sin{\frac{\pi}{18}}.
\end{eqnarray*}
Then all threshold pairs $(\lambda_{\text{th}}^{\text{PSC}},\chi_{
\text{th}}^{\text{PSC}})$ satisfy:
\begin{eqnarray*}
    \lambda_{\text{th}}^{\text{PSC}}&\geq&\frac{1}{2}+\sin \frac{\pi}{18},\\
    \chi_{\text{th}}^{\text{PSC}}&=&\yaqi{\sqrt{\frac{1}{2}\left(1-\frac{2\sin \frac{\pi}{18}}{2\lambda_{\text{th}}^{\text{PSC}}-1}\right)}},
\end{eqnarray*}
\yaqi{which mark the threshold boundary of the PSC protocol (orange dashed line in Fig.~\ref{fig_honeycomb_Linfty}).
This boundary lies mostly below that of the first case (with hybrid-swapping pre-processing), indicating that, for the same resources, the PSC protocol without pre-processing more readily reaches the threshold for entanglement percolation.}

\yaqi{
The inequality $q_{\text{PSC}}> p^{\text{hex}}$ (i.e., $P_{\text{SC}}^{\mathcal{G}_1} > 0$) holds for all $(\lambda,\chi)$ such that $q^{\rm hybrid}> p^{\text{tri}}$ (i.e., $P_{\text{SC}}^{\mathcal{G}_2} > 0$). 
This suggests that hybrid-swapping preprocessing does not improve the performance of the PSC in honeycomb lattices.}

\yaqi{In Fig.~\ref{fig_honeycomb_Linfty}, we show the percolation regions for the PSC protocol without (pink and yellow shaded regions, $P_{\text{SC}}^{\mathcal{G}_{1}}>0$) and with (pink shaded regions, $P_{\text{SC}}^{\mathcal{G}_{2}}>0$) hybrid-swapping pre-processing in an infinite honeycomb lattice.
The region satisfying $q_{\rm PSC}>p^{\rm hex}$ is considerably larger than that satisfying $q^{\rm hybrid}>p^{\rm tri}$, indicating that entanglement percolation is more readily achieved without hybrid-swapping pre-processing.}

{Furthermore, we consider another BSM-based hybrid protocol, which enhances the efficiency of entanglement distribution beyond PSC in the infinite-size hybrid honeycomb lattice.}

Specifically, we consider a swapping protocol {where} each TMSVS is deterministically converted into $|\psi_{\eta}\rangle$ where $\eta=\max\left\{{1}/{2}, 1-\chi^2 \right\}$ with $\chi=\tanh r$, followed by a BSM 
\begin{eqnarray*}
    \mathcal{M}_{\text{BSM}}=\{|\Phi^+\rangle\langle\Phi^+|,|\Phi^-\rangle\langle\Phi^-|,|\Psi^+\rangle\langle\Psi^+|,|\Psi^-\rangle\langle\Psi^-|\}
\end{eqnarray*}
performed on node $R$ where $|\Phi^\pm\rangle=\frac{1}{\sqrt{2}}(|00\rangle\pm|11\rangle)$ and $ |\Psi^\pm\rangle=\frac{1}{\sqrt{2}}(|01\rangle\pm|10\rangle)$.
{Each BSM outcome $j$ occurs with probability $q_j$ and prepares a corresponding conditional pure state $|\phi_j\rangle$ shared between $S$ and $T$.
The resulting ensemble is therefore
$\{q_j,|\phi_j\rangle\}_{j=1}^{4}$, where}
\begin{eqnarray*}
    |\phi_1\rangle&=&\left[\sqrt{\lambda\eta}|00\rangle+\sqrt{(1-\lambda)(1-\eta)}|11\rangle\right]/\sqrt{2q_1},\\
    |\phi_2\rangle&=&\left[\sqrt{(1-\lambda)\eta}|01\rangle+\sqrt{\lambda(1-\eta)}|10\rangle\right]/\sqrt{2q_2},\\
    |\phi_3\rangle&=&\left[\sqrt{\lambda\eta}|00\rangle-\sqrt{(1-\lambda)(1-\eta)}|11\rangle\right]/\sqrt{2q_3},\\
    |\phi_4\rangle&=&\left[\sqrt{(1-\lambda)\eta}|01\rangle-\sqrt{\lambda(1-\eta)}|10\rangle\right]/\sqrt{2q_4},
\end{eqnarray*}
and probabilities $q_1=q_3=\frac{1}{2}
    \left[\lambda\eta+(1-\lambda)(1-\eta)\right]$ and $q_2=q_4=\frac{1}{2}
    \left[(1-\lambda)\eta+\lambda(1-\eta)\right]$, where $\yaqi{\eta}=\max\{1/2,1-\chi^2\}$.
\yaqi{The BSM yields the recorded ensemble $\{q_j,|\phi_j\rangle\}_{j=1}^4$.}
If the outcome is 3 or 4, apply the local unitary transformation $I\otimes Z$. 
The corresponding \yaqi{ensemble-average conditional SCP} is given by
\begin{eqnarray}\label{eq-app-BSM}
    p^{\text{BSM}}=\sum_{j=1}^2 2q_j P_{\text{SCP}}(|\phi_j\rangle)=2\min\{\chi^2,1-\lambda\}.
\end{eqnarray}
\yaqi{Notably, $p^{\text{BSM}}$ is an average over \yaqi{the SCPs of} known conditional pure states with the measurement record retained.}

When taking the \yaqi{BSM-based} hybrid protocol as the swapping protocol on the honeycomb lattice in Sec.~\ref{sec-honeycomb}, \yaqi{all} critical threshold \yaqi{pairs} $(\lambda_{\text{th}}^{\text{BSM}},\chi_{\text{th}}^{\text{BSM}})$ \yaqi{satisfy}
\begin{equation*}
   \yaqi{\lambda_{\text{th}}^{\text{BSM}}\leq 1-\sin{\frac{\pi}{18}},\quad \chi_{\text{th}}^{\text{BSM}}=\sqrt{\sin{\frac{\pi}{18}}}}.
\end{equation*} 
or
\begin{equation*}
   \yaqi{\lambda_{\text{th}}^{\text{BSM}} = 1-\sin{\frac{\pi}{18}},\quad \chi_{\text{th}}^{\text{BSM}}\geq \sqrt{\sin{\frac{\pi}{18}}}.}
\end{equation*} 
Furthermore, \yaqi{when} $\lambda= 1-\sin{({\pi}/{18})}-\delta$ and $\chi=\sqrt{\sin({\pi}/{18})}+\delta$ \yaqi{for $0<\delta\leq 0.03$}, we observe that $q_{\text{PSC}} < p^{\text{hex}}$, indicating that the QN can achieve entanglement percolation with lower initial resource requirements under the BSM-based hybrid protocol than under PSC.
This demonstrates that BSM-based swapping enhances the entanglement distribution efficiency \yaqi{near $(1-\sin{({\pi}/{18})}-\delta,\sqrt{\sin({\pi}/{18})}+\delta)$}.

A natural question arises: can this swapping protocol be applied to general series QNs (Fig.~\ref{fig-hybridQN4})? {We answer this question by analyzing the exponential growth of the conditional-state ensemble generated by successive BSMs. For a single swapping operation, the output can be represented by \yaqi{$\{2q_j,|\phi_j\rangle\}_{j=1}^{2}$}.}
When this process is extended to each node $R_n$ in a series network with $N$ intermediate nodes between $S$ and $T$, every swapping order $s_N$ \yaqi{can yield an ensemble with up to $2^{N}$ conditional branches.}
\yaqi{ Consequently, extending the calculation in Eq.~\eqref{eq-app-BSM} requires tracking an exponentially growing ensemble of conditional output states, making direct exact analytical evaluation increasingly complicated. Therefore, the BSM-based hybrid protocol is not amenable to efficient exact analysis of entanglement distribution in general series networks in its present form.}

{\color{black}{

\section{Illustrative examples}\label{sec-AP-examples}

As discussed in the main text, distance-dependent attenuation is incorporated into the resource-level description by assigning effective CV and DV resource parameters, $\chi_{\rm eff}(L)$ and $\lambda_{\rm eff}(L)$, and substituting them into the network-level formulas according to Eq.~\eqref{substitutions}. This procedure defines an approximate pure-state model for resource estimation, since an initial pure state generally becomes mixed after propagation through a lossy channel. Let $\chi_0$ denote the ratio negativity of each initial CV state and $c_0$ the concurrence of each initial DV state. Assuming that suitable postselection yields pure states of the same forms as the initial resources, we adopt $\chi_{\rm eff}(L)=\chi_0\sqrt{\eta(L)}$ as a phenomenological approximation for CV states and $c_{\rm eff}(L)=c_0\sqrt{\eta(L)}/[1-\lambda_0(1-\eta(L))]$ for DV states, where the concurrence $c_{\rm eff}(L)=2\sqrt{\lambda_{\rm eff}(L)[1-\lambda_{\rm eff}(L)]}$ is obtained from the normalized no-loss branch of a one-sided amplitude-damping channel. Substituting these effective parameters into the pure-state transmission formulas yields conditional SCP values. In the following examples, we do not include the postselection success probabilities. We illustrate this effective-resource model through examples involving elementary links, series networks, and parallel networks.

Consider a fiber link of length $L=100~\mathrm{km}$ with attenuation length $L_{\rm att}=22~\mathrm{km}$, giving $\eta(L)\approx 0.0106$.
We first illustrate how the loss is incorporated into the effective-resource description at the elementary-link level.

For the CV resource with $10~\mathrm{dB}$ squeezing, $\chi_0\approx0.818$. We adopt the resource-level parameterization $\chi_{\rm eff}(L)=\chi_0\sqrt{\eta(L)}$, which gives
$$\chi_{\rm eff}(L)\approx0.0843.$$
For a DV link described by a pure two-qubit
state, the corresponding resource is
parameterized by the Schmidt coefficient
$\lambda_0$. For reference, a Bell-state
fidelity of $F=0.90$ corresponds to
$\lambda=0.80$ within the pure-state model,
according to
$F=1/2+\sqrt{\lambda_0(1-\lambda_0)}$.
This correspondence is illustrative and
does not directly characterize experimental
mixed states.
For $\lambda_0=0.80$, the initial concurrence \yaqi{is} $c_0=0.80$.
We then use the simple resource-level attenuation model $c_{\rm eff}(L)=c_0\sqrt{\eta(L)}/[1-\lambda_0(1-\eta(L))]$.
For $L=100~\mathrm{km}$, the effective concurrence is $c_{\rm eff}(L)
\approx 0.3953$ and the Schmidt coefficient is
$$\lambda_{\rm eff}(L)\approx 0.9593.$$
These effective parameters characterize the individual elementary resources and will be used in the network-level analysis below.

\emph{Simple series network.---}As a simple network example, consider now a general CV-first series hybrid QN with $N$ elementary links in the form $|\psi^{\chi_0}\rangle\text{-}|\psi_{\lambda_0}\rangle\text{-}|\psi^{\chi_0}\rangle\text{-}|\psi_{\lambda_0}\rangle\text{-}\cdots$. 
Then the numbers of CV and DV links are $N_{\rm CV}=\lceil N/2 \rceil$ and $N_{\rm DV}=\lfloor N/2 \rfloor$, respectively.
Let each elementary link have length $L=100~\mathrm{km}$ and attenuation length $L_{\rm att}=22~\mathrm{km}$. Under the effective-link approximation, the network is represented as $|\psi^{\chi_{\rm eff}}\rangle\text{-}|\psi_{\lambda_{\rm eff}}\rangle\text{-}|\psi^{\chi_{\rm eff}}\rangle\text{-}|\psi_{\lambda_{\rm eff}}\rangle\text{-}\cdots.$
For the hybrid-swapping protocol introduced in Sec.~\ref{sec-hybrid_protocol}, the total SCP is
\begin{eqnarray*}
    p_{\rm hybrid}^{\rm seri}=1-\sqrt{1-\left[4\chi_{\rm eff}^2(L)(1-\chi_{\rm eff}^2(L))\right]^{N_{\rm CV}}c_{\rm eff}^{2N_{\rm DV}}(L)}
\end{eqnarray*}
where $c_{\rm eff}(L)=2\sqrt{\lambda_{\rm eff}(1-\lambda_{\rm eff})}$.
In the small-resource regime, 
\begin{eqnarray*}
    p_{\rm hybrid}\sim\left[4\chi_{\rm eff}^2(L)(1-\chi_{\rm eff}^2(L))\right]^{N_{\rm CV}}c_{\rm eff}^{2N_{\rm DV}}(L)/2,
\end{eqnarray*}
showing the exponential decrease of the total SCP with the network size $N$.
The corresponding conditional estimates for several network sizes are shown in Table~\ref{table-resource_estimation}.

\begin{table}[t]
\centering
\caption{Illustrative resource-level conditional estimates for a CV-first series hybrid QN with $100~\mathrm{km}$ elementary links. Here $N$ denotes the number of elementary links and $N+1$ the number of network nodes.}
\label{table-resource_estimation}
\begin{tabular}{c|c|c|c|c}
\hline\hline
$N$ & Nodes & $N_{\rm CV}$ & $N_{\rm DV}$ &
$p^{\rm seri}_{\rm hybrid}\quad / \quad p_{\rm PSC}^{\rm seri}$ \\
\hline
1 & 2 & 1 & 0 &
$1.42\times10^{-2}\quad/\quad1.42\times10^{-2}$ \\
2 & 3 & 1 & 1 &
$2.21\times10^{-3}\quad/\quad1.16\times10^{-3}$ \\
3 & 4 & 2 & 1 &
$6.22\times10^{-5}\quad/\quad1.65\times10^{-5}$ \\
4 & 5 & 2 & 2 &
$9.73\times10^{-6}\quad/\quad1.34\times10^{-6}$ \\
5 & 6 & 3 & 2 &
$2.75\times10^{-7}\quad/\quad1.91\times10^{-8}$ \\
6 & 7 & 3 & 3 &
$4.29\times10^{-8}\quad/\quad1.55\times10^{-9}$ \\
\hline\hline
\end{tabular}
\end{table}

\emph{Simple parallel network.---}Consider a genuinely hybrid parallel QN with two nodes $S$ and $T$ sharing two states $|\psi^{\chi_0}\rangle$ and $|\psi_{\lambda_0}\rangle$. For $L=100~\mathrm{km}$ and $L_{\rm att}=22~\mathrm{km}$ for each elementary link, we use $\chi=\chi_{\rm eff}(L)\approx0.0843$ and $\lambda=\lambda_{\rm eff}(L)\approx0.9593$ in Eq.~\eqref{eq-mu}. The resulting genuine entanglement concentration yields a total SCP of approximately
$$p_{\rm hybrid}^{\rm para}\approx0.0951,$$
which exceeds the SCP under the PSC protocol,
$$p_{\rm PSC}^{\rm para}=1-[1-P_{\rm SCP}(|\psi^{\chi_{\rm eff}}\rangle)][1-P_{\rm SCP}(|\psi_{\lambda_{\rm eff}}\rangle)]\approx 0.0945.$$
Thus, entanglement concentration can improve the total SCP relative to the PSC protocol, albeit only marginally. Moreover, because the entanglement of both resources has already been substantially degraded by the strong attenuation over the $100~\mathrm{km}$ elementary links, the resulting total SCP remains relatively small despite the gain from resource combination.

}
}

\newpage
\clearpage

\bibliography{hybrid_SI}

\end{document}